\documentclass[12pt]{article}

\usepackage{harvard}

\harvardparenthesis{square}
\harvardyearparenthesis{square}

\usepackage{psfrag}

\usepackage{makeidx}
\usepackage{latexsym}
\usepackage{amsthm}
\usepackage{amsmath}
\usepackage{graphicx}

\makeindex

\newtheorem{prop}{Proposition}[section]

\newcommand{\IF}{\mathrm{if}}

\graphicspath{{figure/}}

\title{Time Scaling of Chaotic Systems:\\
	Application to Secure Communications}
\author{Donatello Materassi and Michele Basso\\
        \small Dipartimento di Sistemi e Informatica, Universit\`{a} di Firenze\\
	 \small via S. Marta, 3, I-50139 Firenze (Italy)\\ 
	(\small \texttt{materassi@dsi.unifi.it / basso@dsi.unifi.it})
		}

\date{}
\begin{document}

\maketitle





\begin{abstract}
The paper deals with time-scaling transformations of dynamical systems.
Such scaling functions
 operate
a change of coordinates on the time axis of the system trajectories
preserving its phase portrait.
Exploiting this property, a chaos encryption technique to transmit a binary 
signal through an analog channel
is proposed. The scheme is based on a suitable time-scaling function
which plays the role of a private key. 
The encoded transmitted signal is proved to resist known decryption attacks offering a
secure and reliable communication.
\end{abstract}

\section{Introduction}
In the last decades, encryption schemes that hide messages in
chaotic signals have attracted attention as a tool to
transmit information securely.
The basic principle is to conceal the plaintext message using a chaotic signal and to recover it at the end
of the receiver by means of a synchronization process \cite{PecoraCarrol}.
In literature many techniques have been proposed, but they can be mainly divided into three
different categories: Chaotic Masking, Chaotic Shifting Key and Chaotic Modulation.\\
Chaotic Masking has been the first encryption method introduced in chaotic communications.
Basically, a chaotic signal (the ``mask'') is added to the plaintext to obtain the ciphertext. 
The synchronized receiver is able to recover the plaintext by simply subtracting the ``mask''
\cite{ChaosMask}.\\
In Chaotic Modulation a chaotic signal is modulated by the plaintext and the receiver recovers it
through an ``inversion'' process that obviously depends on the modulation technique
\cite{ChaosModulation}.\\
In a wide sense, Chaotic Shifting Key (CSK) can also be seen as a special case of Chaotic Modulation. 
It allows the transmission  of a binary signal by switching the parameters of two different chaotic systems. 
The receiver determines the bit value according to the success or failure of its synchronization attempt.
Many cryptanalysis tools have been developed in order to evaluate the security of these
schemes and it has been shown that the realization of secure communications based on chaotic encryption 
is still a quite difficult and challenging task \cite{ReturnMap}.\\
In this paper we study classes of  dynamical systems characterized by having
the same phase diagram, but a different time response. This property reveals to be a useful 
countermeasure against known powerful decryption attacks (such as return map attacks).
This suggests the possibility to effectively employ this kind of systems in chaotic communications.
However, analysis of chaotic and, generally, nonlinear systems is quite complex to perform.
This is the reason why an accurate cryptanalysis of communication schemes based on chaos encryption
is difficult to realize and most employed tools are usually numerical simulations.
Nevertheless, this work provides some theoretical results as a support to guarantee security of the system.
The paper is organized as follows. In Section \ref{Time Scaled Systems} the theoretical framework is
described and developed; in Section \ref{Application to Chaos Encryption} we propose a CSK scheme
for secure communications exploiting time scaling functions; Section \ref{Cryptanalytical considerations}
is devoted to some qualitative cryptanalytical considerations and, finally, we provide some simulation
results in Section \ref{Simulation results}.

\section{Time Scaled Systems}\label{Time Scaled Systems} 
Let us consider an autonomous dynamical system described by the differential equation 
\begin{equation}\label{orsys}
	\frac{d}{dt} x = f(x) \qquad\qquad x\in\Re^n, f:\Re^n\rightarrow\Re^n.
\end{equation}
We introduce a modified (``time-scaled'') system 
\begin{equation}\label{modsys}
	\frac{d}{dt} z = f(z) \lambda(z,t) \qquad z\in\Re^n, \lambda:(\Re^n \times \Re)\rightarrow   \Re.
\end{equation}
where $\lambda$ is called ``time scaling function''.
The effect of multiplying all the components of the function $f$ by the same scalar function is 
just to modify the time scale of the original system \cite{SamFur}.
The adoption of time scale functions is a quite common analysis tool in robotics and chemical applications
because it gives some advantages in designing feedback linearizing controllers \cite{SamFur}, \cite{TSobserver} and, under
some non-restrictive conditions, it does not change the stability properties of the system \cite{SamFur}.
However, in this work we are also interested into the fact that time scaling tranformations 
preserve most topological and geometrical properties of the phase diagram. 
For the sake of generality and completeness, we begin considering a time scale transformation depending both
on time $t$ and the state $z$. 
We report and prove some theoretical results which will be helpful for our purposes trying also to give an
extensive overview.
\begin{prop}[Existence of the time scaled solution]\label{PropExistence}
	Let us consider systems (\ref{orsys}) and (\ref{modsys}).
	If there exists a solution 
	$\phi_x(t,x_0)$ of (\ref{orsys}) with initial condition  $x(0)=x_0$
	 and if $\lambda(\cdot,\cdot)$ is ``regular enough'', then
	there exists a scalar function $\tau(t)$ such that $\phi_x(\tau(t),x_0)$ is a solution for
	(\ref{modsys}) with the same initial value $x_0$.
\end{prop}
\begin{proof}
	By hypothesis, the initial value problem
 	\begin{equation}
	  \begin{aligned}
		\frac{dz}{d\tau}= f(z) \\
		z(0)=x_0\\
	  \end{aligned}
 	\end{equation}
	admits a solution $z(\tau)=\phi_x(\tau,x_0)$.
	Consider now
 	\begin{equation}\label{tau(t)}
	  \begin{aligned}
		\frac{d\tau}{dt}= \lambda(z(\tau),t)  \\
		\tau(t_0)=\tau_0.
	  \end{aligned}
 	\end{equation}
	If  $\lambda(z(\tau),t)$ is ``regular enough'',
	there exists a solution $\phi_{\tau}(t,t_0,\tau_0)$. 
	Let $\tau(t):=\phi_{\tau}(t,0,0)$, we can define
	\begin{equation}
		\phi_z(t,x_0):=\phi_x(\tau(t),x_0).
	\end{equation}
	By inspection, $\phi_z(t, x_0)$ is a trajectory of (\ref{modsys}) with $t_0=0$ and
	 $z(t_0)=x_0$.
\end{proof}
We remark that the ``regularity'' required on the time scale function $\lambda$ is needed only to solve 
the Cauchy Problem (\ref{tau(t)}). Actually, it could be sufficient to assume that $\lambda$ is a piecewise
locally Lipschitz function which definitely is not a restrective condition.
However, if $\lambda$ satisfies some additional properties,  some more strict relations beetween
the original and the ``modified'' system can be proved.
\begin{prop}\label{PropEquivalence}
	Under the conditions of Proposition \ref{PropExistence} and assuming that
	$\exists~l,L\in \Re$ such that $ 0< l \leq \lambda(z,t) \leq L $, the phase diagrams of the two
	systems are identical.
\end{prop}
\begin{proof}
	The proof is straightforward. Since $0<l\leq \frac{d\tau}{dt}\leq L$, we can immediately
	conclude that $\tau(t)$ is continuous, increasing monotonic and therefore invertible on $\Re$.
	This means that every trajectory $\{\phi_x(t,x_0)|t \in\Re\}$ is completely mapped into 
	the trajectory $\{\phi_z(t,x_0)|t \in\Re\}$.
\end{proof}
The previous proposition estabilishes a strong bond beetween the two systems. In fact, even though
time responses can be very different, trajectories, attractors and stability properties  of the two systems 
are exactly the same \cite{SamFur}.
It is important to remark this holds for any kind of attractors, including strange attractors.
Such a property will be exploited in the next section to derive a secure communication scheme.

\begin{prop}\label{Proptrajtime}
	Let $v$ be a unitary vector in $\Re^n$. Define $y(t)=v^T x(t)$. Given two time instants 
	$t_1 \leq t_2$, assume that
	$\{t\in [t_1,t_2]| \dot y(t) =0 \} $ is a discrete set.
	Consider also a time scale function
	\begin{equation}
		\lambda(z):= 
		\left\{
			\begin{array}{l}
				\Lambda_0\qquad\IF~\lfloor v^T z/h  \rfloor \mathrm{~is~even}\\
				\Lambda_1\qquad\IF~\lfloor v^T z/h  \rfloor \mathrm{~is~odd}\\ 
			\end{array}
		\right.
	\end{equation}
	where the symbol $\lfloor \cdot \rfloor$ denotes the floor function.
	Assume also that
	\begin{equation}\label{harmonicmean}
		2\left(\frac{1}{\Lambda_0}+\frac{1}{\Lambda_1}\right)^{-1}=1.
	\end{equation}
	Then, under the conditions of Proposition \ref{PropEquivalence}
	\begin{equation}
		\lim_{h \rightarrow 0} \phi_z(t_2-t_1,x_1)=\phi_x(t_2-t_1,x_1).
	\end{equation}
\end{prop}
\begin{proof}

	The function $\lambda$ ``slices'' $\Re^n$ by means of hyperplanes orthogonal to the vector $v$.
	In every ``slice''  (whose width is equal to $h$) the time scale is modified through a constant gain
	which is, alternatively, $\Lambda_0$ or $\Lambda_1$.
	Figure \ref{fig_TSsystem} schematically depicts this situation. 
	\begin{figure}[hbt]
	\psfrag{a}{$\frac{dz}{dt} = f(z)\Lambda_0$}
	\psfrag{b}{$\frac{dz}{dt} = f(z)\Lambda_1$}
	\psfrag{c}{$v$}
	\psfrag{d}{$z(t)$}
	\psfrag{e}{$z_1$}
	\psfrag{f}{$z_2~$}
	\begin{center}
		\includegraphics[width=0.75\textwidth]{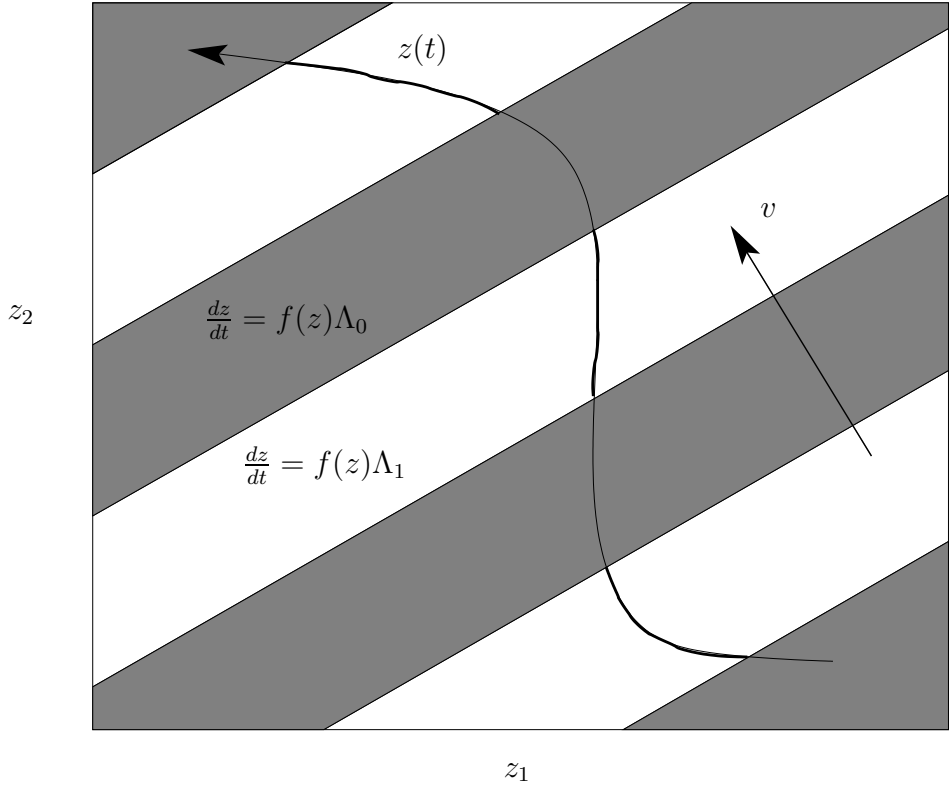}
		\caption{Geometrical interpretation of Proposition
			 \ref{Proptrajtime}. \label{fig_TSsystem}}
	\end{center}
	\end{figure}
	With no loss of generality, we can assume $\Lambda_0\leq\Lambda_1$. 
	Since the harmonic mean (\ref{harmonicmean}) of the two values is one, we
	have $\Lambda_0 \leq 1 \leq \Lambda_1$.
	So, the time scaled system ``moves'' slower than the original one in those regions where
	 $\lambda(z)=\Lambda_0$ and faster where $\lambda(z)=\Lambda_1$.
	However, if those regions are narrow enough (that means $h$ ``small'' enough), the trajectory arc
	 $\widehat{x(t_1) x(t_2)}$  is covered in the same amount of time by both the original and the scaled system.
\end{proof}

\section{Application to Chaos Encryption}\label{Application to Chaos Encryption}
\begin{figure}[hbt]
	\psfrag{A}{$A_i$}
	\psfrag{B}{$B_i$}
  \begin{center}
     \includegraphics[width=0.75\textwidth]{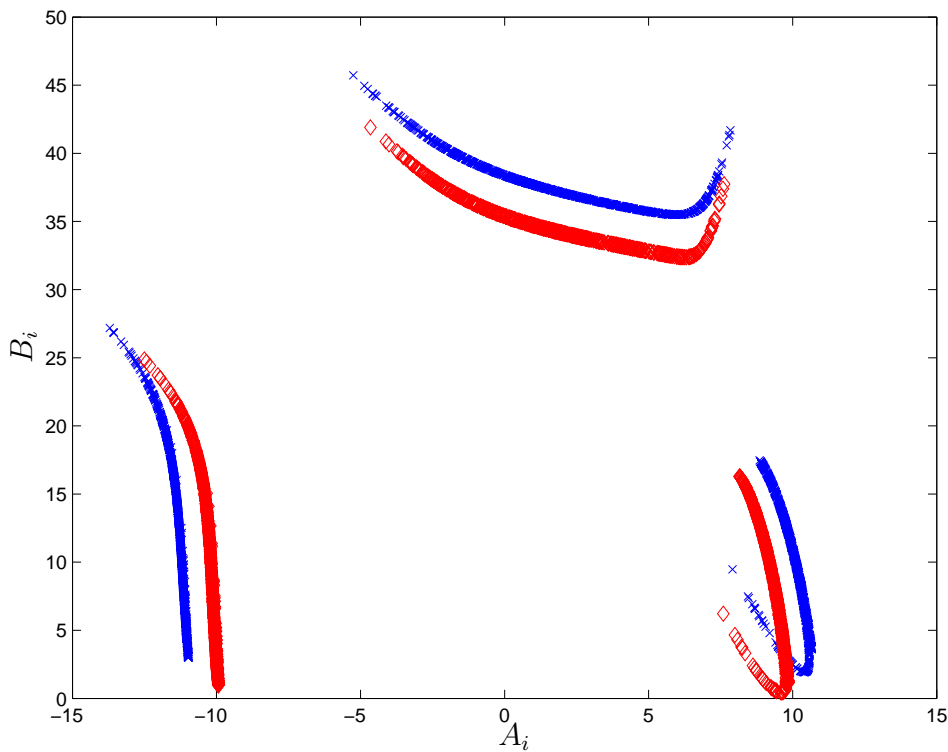}
     \caption{Two different values of the parameter $\theta$ in a CSK scheme can produce 
			two distinguishable return maps.  \label{fig_RMAttack}}
  \end{center}
\end{figure}

In this section we will show how two time-scaled chaotic systems can be employed to send a binary signal
in an encrypted manner.
In particular, we will use two time-scaled chaotic systems in a chaotic shifting key (CSK) scheme
showing how some security issues can be solved.\\
In a basic CSK scheme, the plaintext is encoded as a sequence of chaotic signals produced by
one of two different chaotic systems according to the bit value. The receiver decodes the ciphertext
through a simple on-off synchronization process.
Consider the following system used as the sender
\begin{equation}\label{genericsender}
 	\dot x = f(x,s,\theta(m))
\end{equation}
where $s$ is a scalar component of the state $x$ and
$\theta(m)$ is a vector of parameters modulated by a binary plain signal $m(t)$
\begin{equation}\label{CSKSwtichtheta}
	\theta=\theta(m)=
	\left\{
 		\begin{array}{l}
			\theta_0 \qquad \IF~m=0\\
			\theta_1 \qquad \IF~m=1.
		\end{array}
	\right.
\end{equation}
Both $\theta_0$ and $\theta_1$ must have been suitably chosen to generate a chaotic regime in (\ref{genericsender}).
To transmit $m(t)$, the signal $s(t)$ is sent out in order to cause synchronization to the receiver
\begin{equation}
	\dot z = f(z,s,\theta_1).
\end{equation}
If synchronization is achieved the bit is revealed to be $1$,
 while, if there is no synchronization, the bit is concluded to be $0$.
The security of the system is based on the fact that an intruder would observe only the 
``apparently meaningless'' 
chaotic signal $s(t)$ and should not be able to achieve synchronization without an accurate knowledge of the parameters
$\theta_0$ and $\theta_1$ which play the role of a private key.
Nevertheless, this basic scheme has been proved to be very vulnerable to Return Map attacks \cite{ReturnMap}.
In fact, assuming that $x_i$ and $X_i$ are the $i-th$ minima and maxima, respectively, of $s$, we define
the following variables $A_i:=X_i+x_i$ and $B_i:=X_i-x_i$. The plot of $B_i$ as a function of $A_i$ is
called Return Map (RM) of the signal $s$. 
The RM is topologically equivalent to the peak to peak dynamics plot described in \cite{CandatenRinaldi}.
An intruder, observing the encrypted signal $s(t)$, can  easily reconstruct the RM with no knowledge
of the parameters $\theta_0$ or $\theta_1$. In fact, if the two RMs of system (\ref{genericsender}) when 
$\theta=\theta_0$ and when $\theta=\theta_1$ are ``well distinguishable'', it is possible to unmask 
the concealed bit simply checking which map the transmitter is currently tuned on.
An example of a  RM reconstruction is reported in Figure \ref{fig_RMAttack}. In this case the obtained RM
shows two evident branches, one is associated to $\theta_0$ and the other one to $\theta_1$ . 
The intruder, by simply checking what branch the transmitted signal is currently associated, can easily
recover the plaintext sampled at every peak.
Of course, the presence of channel noise would disturb the attacker because he would obtain
blurred maps. At the same time, we have to consider that noise would produce negative consequences on
the receiver synchronization process, too.
Many countermeasures have been considered in order to resist RM attacks (such as \cite{BuWang},
\cite{Paniyandi}), and many of  them have been broken, as well \cite{ReturnMap}.
In \cite{XuChee} a very interesting approach has been proposed where the parameter $\theta$ still 
switches according to the plain-text bit value, but some more additional random switches are introduced
to confuse a possible intruder. In this case, the switching policy (\ref{CSKSwtichtheta}) 
is extended to the following form
\begin{equation}
	\theta=\theta(x,m)
\end{equation}
The additional pseudo-random switches occur according to one or more
state variables (that are not transmitted), such that they do not 
disturb the receiver when it is in a synchronized
condition. Conversely,
this is novelty of the idea, an attacker trying to reconstruct the RM will be frustrated by the presence of many
switches s/he is not able to predict because they are intrinsically related to the knowledge of the system
structure.
However, it is our opinion that the pseudo-random switch approach, in its general form described
in \cite{XuChee}, is still structurally vulnerable to RM attacks, if some precautions are not taken.
Moreover, there are no theoretical results to guarantee that the reconstructed RM can not actually be 
exploited by an attacker.

\subsection{Time Scale CSK scheme}
The Time Scale CSK (TS-CSK) communication scheme we propose is partially inspired by \cite{XuChee}, but it adopts time scaling functions to prevent the system to be  broken by standard return map attacks.
The transmitter and receiver have the following structure
\begin{equation}\label{TS-CSKgeneric}
	\begin{array}{l}
	\dot x = f(x,s)\lambda(x,m)\\
	\dot z = f(z,s)\lambda(z,1)
	\end{array}
\end{equation}
where $s$ is one of the state components which is being transmitted and 
$\lambda(x,m)$ is the strictly positive time scaling function
\begin{equation}\label{TS-CSKspecific}
	\lambda(x,m)=
		\left\{
			\begin{array}{l}
				\Lambda_{m\phantom{-1}}\qquad\IF~\lfloor v^T z/h  \rfloor
					\mathrm{~is~even}\\
				\Lambda_{1-m}\qquad\IF~\lfloor v^T z/h \rfloor 
					\mathrm{~is~odd}\\ 
			\end{array}
		\right\}
\end{equation}
such that $\Lambda_0$, $\Lambda_1$, $v$ and $h$ are fixed parameters chosen 
to meet condition of Proposition \ref{Proptrajtime} which play .the role of the
encryption key.
In this case, the function $\lambda$ defines a time-scale analogous to the time-scale
described in Proposition \ref{Proptrajtime} where the transmitted bit $m$ simply
inverts the roles of $\Lambda_0$ and $\Lambda_1$.
This choice of $\lambda$ is very demonstrative and it is motivated by its simplicity and by the
theoretical results proved in the previous section.
This peculiar structure allows to carry out some qualitative cryptanalytical considerations.
Of course, a more sophisticated choice could bring to better results in terms of security,
synchronization time and practical realization.
%

\section{TS-CSK Cryptanalysis}\label{Cryptanalytical considerations}
In this section we will analyze how the proposed TS-CSK communication scheme can resist most common decryption attacks.
This study does not intend to be an exhaustive cryptanalysis since this can only be be accomplished describing the exact typology of attacks (known plaintext, known
ciphertext, etc...).
It is our aim to report only some qualitative considerations supported, when possible, by theoretical results. 

\subsection{Return Map attack}
The communication system (\ref{TS-CSKgeneric}) 
is intrinsecally robust against a return map attack.
In fact, from Proposition \ref{PropEquivalence}, it is obvious  that,
under practically non-restrictive conditions on
$f$ and $\lambda$, neither the modulation $m$ nor the pseudo-random switches
modify the phase portrait of the original non-timed scaled system
\begin{equation}\label{againtheoriginalsystem}
	\dot x = f(x,s).
\end{equation}
Therefore, the RMs of the TS-CSK scheme (\ref{TS-CSKgeneric}) 
and system (\ref{againtheoriginalsystem})
are exactly the same regardless of the transmitted bit $m(t)$.

\subsection{Return Time Map attack}
Assuming that $X_i$ is the $i-th$ local maximum of $s(t)$ and $t_i$ is the relative time instant
when it occurs, we define the Return Time Map (RTM) as  the plot of $t_{i+1}-t_i$ versus  $X_i$. 
With no conceptual differences, the RTM could have been defined using the minima of $s(t)$ 
(or both minima and maxima) \cite{CandatenRinaldi}.
The effect of the function $\lambda$ is to ``speed up'' the system when $\lambda>1$ and to ``slow down''
it when  $\lambda<1$. It is intuitive that a wrong choice of the function $\lambda$ could lead to a breakable
system using a RTM attack \cite{CandatenRinaldi}.


By Proposition \ref{Proptrajtime}  it is immediate to conclude that if the number of time scaling
switches is ``dense'' enough (with a proper choice of $\lambda$), then the RTM plot is not ``significantly''
modified by the time scaling.
Of course, this is just a theoretical consideration, since the choice of a too small value for $h$ would negate
the possibility of a physical realization of such communication devices.
However, Proposition \ref{Proptrajtime} confirms the intuition suggesting
small values for $h$ in order to increase the security of the system.

\subsection{Switch detection}
In a CSK scheme, an intruder eavesdropping the communication could detect the changes of bit values
by simply detecting discontinuities in the first derivative of the encoded signal $s(t)$.
This is the reason why the adoption of a continuous function $\lambda$ is deprecated. In fact, it would not 
create discontinuous false-switches analogous to the informative bit switches in order to
confuse the intruder \cite{XuChee}.
If false-switches are frequent enough in time and do not depend on the drive signal
it will be difficult to distinguish which ones will be informative.
Again, the choice of small values for $h$ seems to lead to increased security.

\subsection{Brute Force attack}
In a security analysis, it must be assumed that the intruder knows everything about the communication
system structure, encryption method, physical characteristics (channel noise power spectrum etc...) but
the encryption key. Nevertheless, the encryption key can be guessed, so it is important 
the key space is large enough in order to make such a guess the most difficult it is possible.
A common problem in chaos communication is the fact that physical systems show a chaotic behaviour
only in a very restricted range of their parameters limiting the choice of the key.
The proposed approach, in its general form (\ref{TS-CSKgeneric}), overcomes this problem since there is 
no particular restriction on the function $\lambda$ once it is assumed it is discontinuous and strictly positive.

\section{Example}\label{Simulation results}
\begin{figure}[hbt]
	\psfrag{A}{$A_i$}
	\psfrag{B}{$B_i$}
  \begin{center}
     \includegraphics[width=0.75\textwidth]{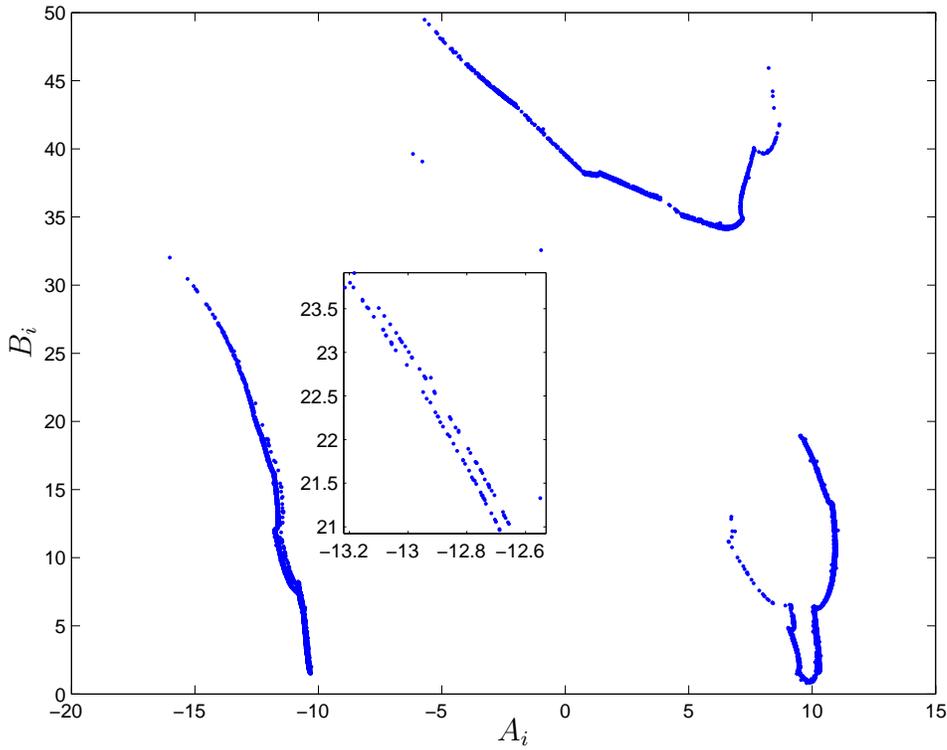}
     \caption{  The Return Map of the PRS-CSK scheme (\ref{PRS-CSK}), assuming no noise acting on che
			 channel, still presents two distinct branches even adopting the pseudo-random switching CSK.
			  \label{fig_RMswitchattack}}
  \end{center}
\end{figure}

The communication system Pseudo Random Switch CSK (PRS-CSK) described in \cite{XuChee} 
exploits the well-known synchronization properties of two identical Lorenz models \cite{PecoraCarrol}
\begin{equation}\label{PRS-CSK}
		\begin{array}{l}
			\dot x_1 = \sigma(x_2 -x_1)\\
			\dot x_2 = (\beta-x_3)x_1 - x_2\\
			\dot x_3 = x_1 x_2 - \rho x_3\\
			~\\
			s=x_1
		\end{array}
\end{equation}
\begin{equation}\nonumber
		\begin{array}{l}
			\dot z_1 = \sigma(z_2 -z_1)\\
			\dot z_2 = (\beta_0-z_3)s - z_2\\
			\dot z_3 = z_2 s - \rho z_3 
		\end{array}
\end{equation}
where $\theta(x,m)=[\sigma, \beta(m), \rho(x)]$. 
The parameter $\beta$ is modulated the plain-text signal $m(t)$ 
\begin{equation}
	\beta(m)=
		\left\{
			\begin{array}{l}
				\beta_0 \quad\IF~ m=0\\
				\beta_1 \quad\IF~ m=1
		\end{array}
			\right.
\end{equation}
while the parameter $\rho$ generates the false switching events
\begin{equation}
	\rho=\rho(x)=
		\left\{
			\begin{array}{l}
				b_1 \quad\IF~ a_1<x_2<a_2\\
				b_2 \quad\IF~ a_3<x_2<a_4\\
				b_3 \quad\IF~ a_5<x_2<a_6\\
				b_4 \quad\IF~ a_7<x_2<a_8\\
				b_5 \quad\mathrm{otherwise}
			\end{array}
		\right.
\end{equation}
In \cite{XuChee} the numerical choice of the variables was 
\begin{equation}
\begin{array}{l}
	\sigma=10\\
	\beta_0 = 60.5\\
	\beta_1 = 60\\
	a=[0, 5, 31, 32, 23, 28, 10, 16]\\
	b=[10/3, 8/3, 2/3, 5/3, 2].
\end{array}
\end{equation}
The transmission of $100$ alternate bits (worst case scenario for the attacker)
(as in \cite{XuChee}) has been simulated 
assuming the absence of channel noise using a standard ODE45 integration method with a relative 
tolerance of $1e-12$  in order to obtain a low numerical noise.
The result of the RM reconstructed by an eventual attacker is depicted in Figure \ref{fig_RMswitchattack}.
As it is shown, in the RM two branches are still distinguishible even though they are definetely close.
However, the close distance of the two branches is mainly related to the fact that the two values 
$\beta_0$ and $\beta_1$ are pretty similar.
In a practical situation, the two branches would be very likely undistinguishable because of channel noise, 
but, as we have previously stressed, the presence of noise may disturb the receiver synchronization as well.
A theoretical guarantee that a RM attack is really ineffective would be desirable from
the security point of view.

\begin{figure}[hbt]
	\psfrag{A}{$A_i$}
	\psfrag{B}{$B_i$}
  \begin{center}
     \includegraphics[width=0.75\textwidth]{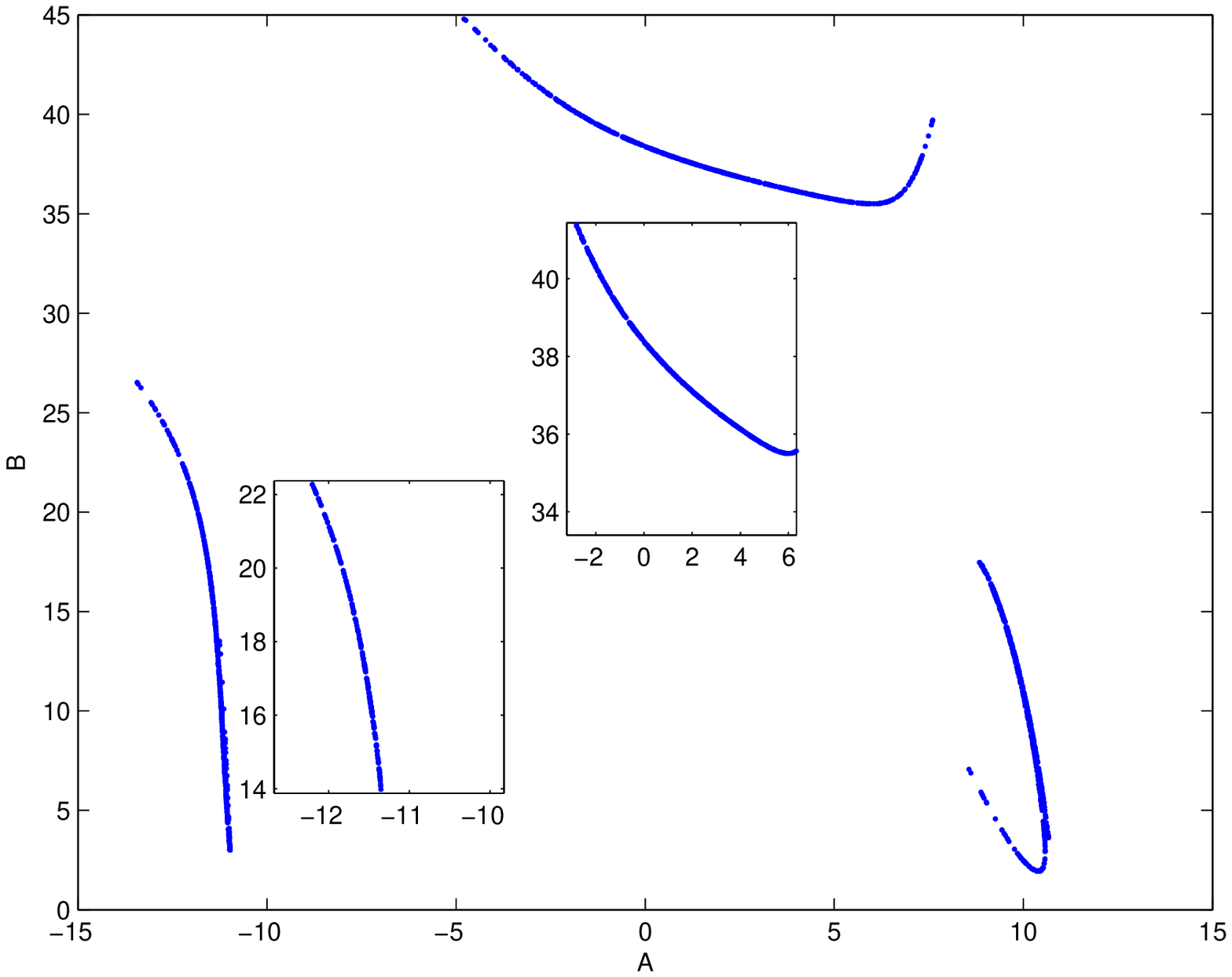}
     \caption{TS-CSK: Return Map attack is completely countered since the RM of the
	 the sender dynamics in (\ref{TS-CSK}) does not change according to the bit
	 value. \label{fig_RMswitchTSattack}}
  \end{center}
\end{figure}

\begin{figure}[hbt]
  \begin{center}
     \includegraphics[width=0.75\textwidth]{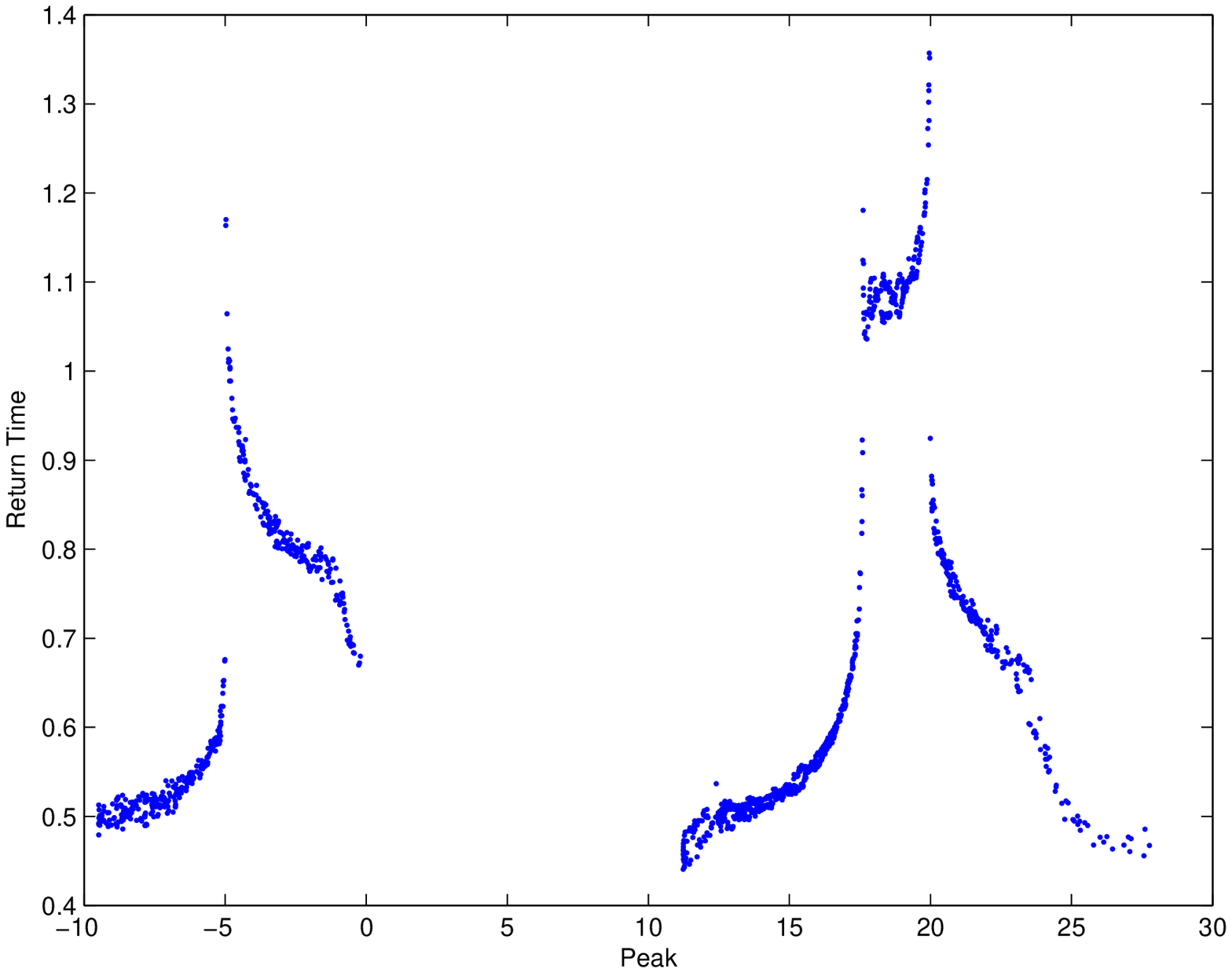}
     \caption{  TS-CSK:
			Return Time attack originates  a sparse  map which makes it
			difficult to crack the system. \label{fig_RTMattack7}}
  \end{center}
\end{figure}

\begin{figure}[hbt]
  \begin{center}
     \includegraphics[width=0.75\textwidth]{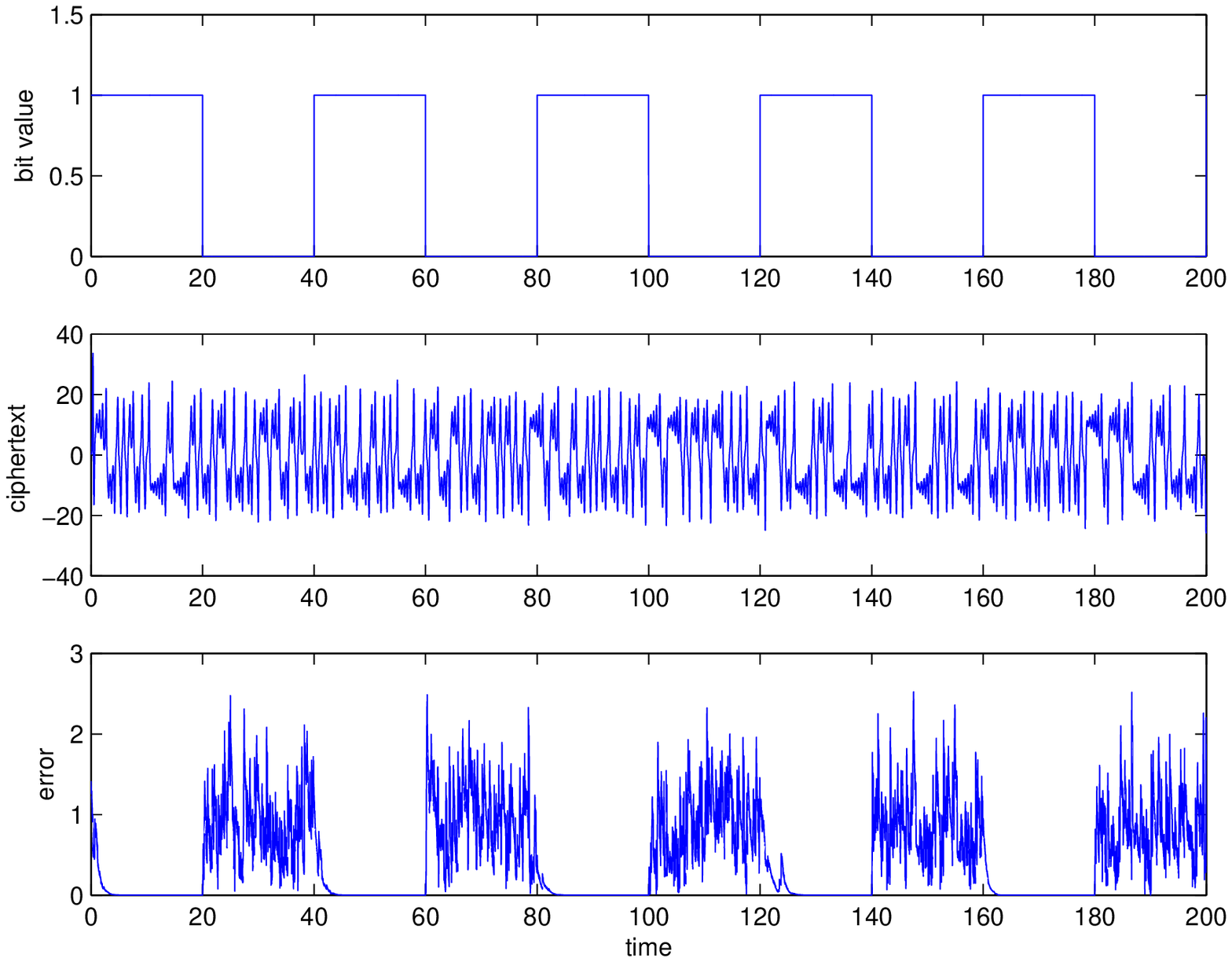}
     \caption{TS-CSK: Transmission of ten alternate bits, the ciphertext and the relative
	 synchronization error
			at the receiver end.
		 \label{fig_bittransmission7}}
  \end{center}
\end{figure}

As a comparison, we employ the same Lorenz model in a TS-CSK communication scheme
\begin{equation}
		\begin{array}{l}\nonumber
			\dot x_1 = [\sigma(x_2 -x_1)]\lambda(x,m)\\
			\dot x_2 = [(\beta-x_3)x_1 - x_2]\lambda(x,m)\\
			\dot x_3 = [x_1 x_2 - \rho x_3]\lambda(x,m)\\
		\end{array}
\end{equation}
\begin{equation}\label{TS-CSK}
			s=x_1
\end{equation}
\begin{equation}
		\begin{array}{l} \nonumber
			\dot z_1 = [\sigma(z_2 -z_1)]\lambda(z,1)\\
			\dot z_2 = [(\beta-z_3)s - z_2]\lambda(z,1)\\
			\dot z_3 = [z_2 s - \rho z_3]\lambda(z,1) 
		\end{array}
\end{equation}
with $\sigma=10$, $\beta=60$, $\rho=2$
and using $\lambda(x,m)$ in (\ref{TS-CSKspecific}) with the private 
encryption/decryption key
\begin{align*}
	v=[0, 1, 0]^T\\
	h=2\\
	\Lambda_0=15/16\\
	\Lambda_1=15/14.
\end{align*}
\begin{figure}
	\begin{center}
\psfrag{x1}{$x_1$}
\psfrag{x2}{$x_2$}
\includegraphics[width=0.75\textwidth]{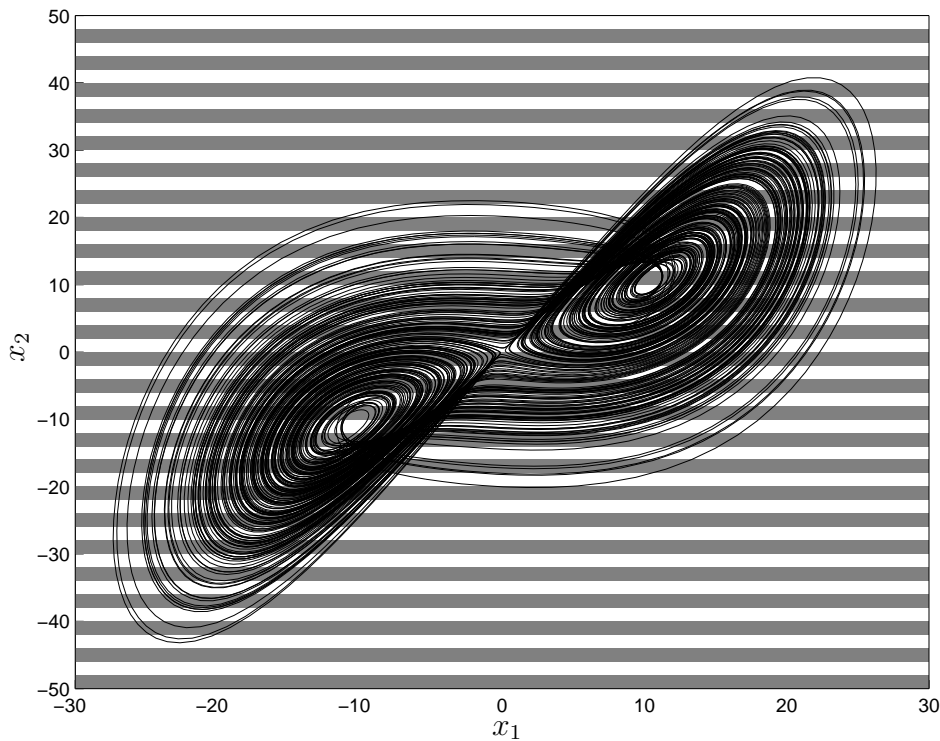}
\caption{Projection of the Lorenz attractor on the plane $x_1$-$x_2$ for the TS-CSK scheme.
 \label{TS-CSKphaseportrait}}
	\end{center}
\end{figure}
Figure \ref{TS-CSKphaseportrait} shows the Lorenz attractor projected on the 
plane $x_1$-$x_2$ along with the time-scale policy. Given the bit $m$ to be  transmitted,
in the white stripes the system evolves with the time-scale factor $\Lambda_m$ 
while in the gray ones the time-scale factor $\Lambda_{1-m}$ is used. 
System (\ref{TS-CSK}) has been simulated with the transmission of $100$ alternate bits (ODE45 solver with relative precision equal to $1e-12$).
The corresponding RM and RTM obtained by an eventual intruder are reported in Figures \ref{fig_RMswitchTSattack}
and \ref{fig_RTMattack7}, respectively. 
Return Map Attack is completely countered since the RM of the
	 the sender dynamics does not change according to the bit value.
Moreover, a RTM based attack does not seem to be so effective because the map is sparse enough
not to reveal the presence of distinct branches (we are implicitly assuming that the integration error in
the simulation procedure can be safely neglected).
In order to show that the proposed scheme can provide a secure and reliable communication, Figure
\ref{fig_bittransmission7} depicts the simulation results of the decryption phase at the receiver.
Ten alternate bits have been encoded and the right synchronizations/desynchronizations occur 
very promptly.

\section{Conclusion}
An encryption scheme to send digital data through an analog channel has been introduced by
exploiting a class of time scaling functions.
It has been proved that such a scheme is intrinsecally secure against simple return map attacks.
Some theoretical results show that return time map attacks should not be so effective, too,
if the choice of the time-scaling function satisfies some requirements.
Switch detection beetween bits in the plaintext is also made difficult by using a pseudo-random
false-switching technique.
Simulations show effectiveness of the proposed communication scheme with respect to
known decryption attacks.

\bibliographystyle{ijbc}
\bibliography{biblio}

%
%
%

\end{document}